\documentclass{amsproc}

\usepackage{amssymb}

\newtheorem{theorem}{Theorem}[section]

\theoremstyle{definition}

\theoremstyle{remark}

\numberwithin{equation}{section}

\begin{document}

\title{Higher spin AdS$_3$ holography and superstring theory}


\author{Thomas Creutzig}
\address{Department of Mathematical and Statistical Sciences, 632 CAB, University of Alberta, Edmonton, Alberta T6G 2G1, Canada}
\email{creutzig@ualberta.ca}
\thanks{The work of TC is supported by NSERC grant number RES0019997.}

\author{Yasuaki Hikida}
\address{Department of Physics, Rikkyo University, 3-34-1 Nishi-Ikebukuro, Toshima, Tokyo 171-8501, Japan}
\email{hikida@rikkyo.ac.jp}
\thanks{The work of YH is supported by JSPS KAKENHI Grant Number 24740170.}

\author{Peter B. R\o nne}
\address{University of Luxembourg, Mathematics Research Unit, FSTC,
Campus Kirchberg, 6, rue Coudenhove-Kalergi, L-1359 Luxembourg-Kirchberg, Luxembourg}
\email{peter.roenne@uni.lu}
\thanks{The work of PBR is funded by AFR grant 3971664 from Fonds National de la Recherche, Luxembourg, and partial support by the Internal Research Project GEOMQ11 (Martin Schlichenmaier), University
of Luxembourg, is also acknowledged. }


\subjclass[2010]{Primary 81T40, 83E30}

\date{\today}

\begin{abstract}
It has been believed for a long time that the tensionless limit of superstring theory can be described by a higher spin gauge theory. Recently, a concrete realization of this idea was proposed via 3d Aharony-Bergman-Jafferis (ABJ) theory with the help of holographic duality. In this note we review our work on finding a similar relation involving 2d coset type models. We start by proposing and examining holographic dualities between 3d higher spin gauge theories with matrix valued fields and the large $N$ limit of 2d coset type models. After that we discuss possible relations to superstring theory with emphasis on the role of the matrix form of the higher spin fields and the extended supersymmetry.

\end{abstract}

\maketitle


\section{Introduction}

The gauge theory of higher spin fields can be introduced as a natural extension of the electromagnetic theory with a spin-1 field and gravity theory described by a spin-2 field.
Superstring theory includes a large spectrum of massive higher spin states and it is believed that the tensionless limit of superstring theory can be described by a higher spin gauge theory.
Moreover, several examples of AdS/CFT dualities involving higher spin gauge theories are known, and while these dualities are much simpler than the full superstring dualities they do share important non-trivial features.

The most famous example of a non-trivial higher spin gauge theory is given by the Vasiliev theory \cite{Vasiliev:2003ev}. It was proposed that 4d Vasiliev theory is dual to the 3d O$(N)$ vector model \cite{Klebanov:2002ja} (see also \cite{Sezgin:2002rt}) and this proposal was confirmed by examining the correlation functions \cite{Giombi:2009wh,Giombi:2010vg}. The role of higher spin symmetry in the correspondence was clarified in \cite{Maldacena:2011jn,Maldacena:2012sf}. Further, it is possible to extend the Vasiliev theory to include matrix valued fields i.e. to associate Chan-Paton (CP) factors. Recently, it was argued in \cite{Chang:2012kt} that the 4d extended Vasiliev theory with CP factors is dual to the 3d Aharony-Bergman-Jafferis (ABJ) theory. Since the ABJ theory is known to be dual to the superstring theory on AdS$_3 \times \mathbb{C}$P$^3$ \cite{Aharony:2008ug,Aharony:2008gk}, the duality in \cite{Chang:2012kt} suggests a non-trivial relation  (called as ABJ triality) between higher spin theory, ABJ theory and superstring theory. In other words, through the AdS/CFT correspondence, it becomes possible to examine superstring theory in terms of gauge theory with a large amount of higher spin symmetry.

The lower dimensional version of the proposal by  \cite{Klebanov:2002ja} was introduced in \cite{Gaberdiel:2010pz} (see \cite{Gaberdiel:2012uj} for a review), and the claim is that the 3d Vasiliev theory in \cite{Prokushkin:1998bq} is dual to a 2d large $N$ minimal model. This proposal is motivated by the enhancement of the asymptotic symmetry in 3d higher spin gauge theory found in \cite{Henneaux:2010xg,Campoleoni:2010zq,Campoleoni:2011hg}. There are several generalizations of this proposal. A truncated version was proposed in \cite{Ahn:2011pv,Gaberdiel:2011nt}.
Moreover, supersymmetric versions were introduced in \cite{Creutzig:2011fe,Creutzig:2012ar,Beccaria:2013wqa} and several supporting checks can be found, e.g., in \cite{Candu:2012jq,Hanaki:2012yf,Henneaux:2012ny,Creutzig:2012xb,Moradi:2012xd}. It is also possible to extend the 3d Vasiliev theory to include matrix valued fields \cite{Prokushkin:1998bq}. Therefore, it is natural to expect that the AdS/CFT correspondence with the 3d extended Vasiliev theory with CP factors leads to a non-trivial correspondence between higher spin gauge theory and superstring theory as in \cite{Chang:2012kt}. We expect to obtain a deeper understanding about such trialities  by studying the lower dimensional version since, in general, lower dimensional theories are more tractable than higher dimensional ones. In this note we would like to review our works on this subject in \cite{Creutzig:2013tja,Creutzig:2014ula}. Similar works may be found in \cite{Gaberdiel:2013vva,Gaberdiel:2014yla,Beccaria:2014jra,Gaberdiel:2014cha,Candu:2014yva}.

The rest of this note is organized as follows;
In the next section we review the ABJ triality in \cite{Chang:2012kt} and explain why it is important to introduce  CP factors to the higher spin fields.
In section \ref{CP} we propose a duality between 3d higher spin gauge theory with CP factors and a 2d coset type model, and give several supporting arguments for the proposal.
In section \ref{N=3} we slightly generalize the duality to accommodate extended supersymmetry, and discuss the relation to superstring theory by making use of the $\mathcal{N}=3$ superconformal symmetry of the 2d coset  type model.
In section \ref{conclusion} we conclude this note.

\section{A review of ABJ triality}
\label{ABJ}

In order to explain the ideas on the ABJ triality in \cite{Chang:2012kt}, let us start form
the original proposal by Klebanov and Polyakov \cite{Klebanov:2002ja}, where the minimal Vasiliev theory on AdS$_4$ is dual to the 3d O$(N)$ vector model. The Vasiliev theory includes totally symmetric tensor fields
$ \varphi_{\mu_1 \ldots \mu_s}$, which transform under the gauge transformation as
\begin{align}
 \varphi_{\mu_1 \ldots \mu_s} \sim \varphi_{\mu_1 \ldots \mu_s} + \partial_{( \mu_1} \xi_{\mu_2 \ldots \mu_s )} \, .
\end{align}
Here $\xi_{\mu_1 \ldots \mu_{s-1}}$ are gauge parameters and the parenthesis means symmetrization of the indices.
In the minimally truncated case, the theory includes a  gauge field for each even spin $s=2,4,6,\ldots$. The dual theory is proposed to be 3d O$(N)$ vector model, which consists of $N$ free bosons $h_i$ $(i=1,2,\ldots ,N)$ in the vector representation of the O$(N)$ global symmetry.
We need to take the large $N$ limit to relate to the classical theory of higher spin fields.
We also assign an O$(N)$ singlet condition to the operators, which may then be given by bilinears of $h_i$ such as $h_i h^i$. The operators dual to the higher spin gauge fields are then constructed as
\begin{align}
 J_{\mu_1 \ldots \mu_s} = h_i \partial_{( \mu_1} \cdots \partial_{\mu_s )} h^i + \cdots
\end{align}
 by the action of derivatives.
 
In \cite{Chang:2012kt} they extended this duality in order to relate to superstring theory. There are several crucial points such as the introduction of supersymmetry, deformation from the free theory, and so on. Among them, here we would like to focus on the CP factor for the higher spin gauge fields. In the Vasiliev theory, it is not so difficult to extend the theory to
include matrix valued fields. Field equations in the Vasiliev theory are given in terms of a non-commutative $*$-product. Now we require the fields to take $M \times M$ matrix values such as $ [\varphi_{\mu_1 \ldots \mu_s}]^\alpha_{~\beta} $ with $\alpha ,\beta = 1,2, \ldots, M$. Due to the replacement, we need to change the $*$-multiplication by including also the multiplication in the matrix algebra. Even after the changes, we can use the same field equations since the $*$-product already has a non-abelian nature.

A version of the 4d extended Vasiliev theory is proposed in \cite{Chang:2012kt} to be dual to the ABJ theory, which is a 3d $\text{U}(N) \times \text{U}(M)$ Chern-Simons matter theory. The theory includes bi-fundamental matter, i.e. fields $A_i^\alpha, B_\beta^j$ in the bi-fundamental representation of $\text{U}(N) \times \text{U}(M)$. We need to take a large $N$ limit to get the relation to the classical higher spin theory. As in the  case without CP factor, let us assign a U$(N)$ invariant condition. Then we can construct higher spin currents in terms of bi-linears of bi-fundamentals as
\begin{align}
 [J_{\mu_1 \ldots \mu_2}]^\alpha_{~ \beta} = A_i^\alpha  \partial_{( \mu_1} \cdots \partial_{\mu_s )} B^i_\beta + \cdots  \, .
\end{align}
In this way we can construct $M \times M$ matrix valued currents dual to $M \times M$ matrix valued higher spin fields. In order to have operators dual to string states, we have to also assign a U$(M)$ invariant condition as well.
The U$(M)$ singlets are given by single trace operators in the form of $\text{tr} \, ( ABAB \cdots AB ) $. Since the bi-linear $AB$ is supposed to correspond to a single-particle state in higher spin theory, this implies that a generic string corresponds to a multi-particle state of higher spin fields in the singlet of the U$(M)$ symmetry for the CP factor. This is a lesson we obtained from the ABJ triality in \cite{Chang:2012kt}, and we shall utilize it in order to construct a lower dimensional analogue.

\section{Higher spin AdS$_3$ holography with CP factor}
\label{CP}

We now try to find an AdS$_3$ version of the ABJ triality.
This should relate higher spin theory on AdS$_3$, 2d CFT and superstring theory
 on AdS$_3 \times $M$_7$. Here M$_7$ represents a 7d manifold.
{}From the analysis of the ABJ triality, we should extend the 3d Vasiliev theory such that the theory includes $M \times M$ matrix valued fields, which was actually constructed in \cite{Prokushkin:1998bq}. First we propose which model is the 2d CFT dual to the 3d Vasiliev theory with CP factors, and check the proposal. Then we will  discuss the relation to superstring theory.

We consider the 3d Vasiliev theory with $M \times M$ matrix valued fields and also with $\mathcal{N}=2$ supersymmetry.  The theory includes massive matter fields along with higher spin gauge fields. The gauge algebra is a supersymmetric higher spin algebra denoted by shs$_M [\lambda]$, and the masses of the matter fields are also parametrized by the same parameter $\lambda$.
The proposal is that the dual theory is given by the following coset \cite{Creutzig:2013tja}
\begin{align}
  \frac{\text{su}(N+M)_k \oplus \text{so}(2NM)_1}{\text{su}(N)_{k+M} \oplus \text{u}(1)_\kappa}
  \label{coset}
\end{align}
with $\kappa = NM (N+M) (N+M+k)$.
In order to relate to the classical higher spin theory, we take a large $N$ limit where $N,k\to \infty$ while we keep $M$ finite as well as the 't Hooft parameter
\begin{align}\label{}
    \lambda = \frac{N}{N + M + k} \, .
\end{align}
This 't Hooft parameter is identified with $\lambda$ appearing in the dual higher spin theory. Moreover, $M$ is set to be the same as the size of CP factor.
For $M=2$ our duality reduces to the one in \cite{Gaberdiel:2013vva} obtained independently.

There are several results that support our conjecture. First of all, we can see that the proposal is a natural extension of the previously known duality without CP factor. Indeed, the coset \eqref{coset} with $M=1$ reduces to the coset used in the duality of \cite{Creutzig:2011fe}, which is an $\mathcal{N}=2$ supersymmetric extension of the original proposal in \cite{Gaberdiel:2010pz}. Moreover, in the  limit of large level $k \to \infty$ (or $\lambda \to 0$ in terms of the 't Hooft parameter) the coset can be shown to reduce to a free system with bi-fundamentals.
The group manifold SU$(N+M)$ may be described by an $(N + M) \times (N \times M) $ matrix as
\begin{align}
 \begin{pmatrix}
 A & B\\
  C & D
 \end{pmatrix} \, .
\end{align}
Ignoring the U$(1)$ factor, $A$ corresponds to the gauge factor SU$(N)$ in the denominator of the coset \eqref{coset}, while $D$ represents SU$(M)$ symmetry which can be shown to decouple in the limit. The other blocks $B,C$ transform as bi-fundamental representations under the $\text{SU}(N) \times \text{SU}(M)$ transformation.
The limit of large level $k$ corresponds to the small curvature limit of the coset manifold, and the bi-fundamentals become free boson fields. Therefore, we can apply the arguments for the ABJ triality in section \ref{ABJ}. {}From the bilinears of the bi-fundamentals,
we can construct higher spin currents of $M \times M$ matrix form, and we can see that they are dual to the higher spin gauge fields with U$(M)$ CP factor at the parameter value $\lambda = 0$.

Even with generic $M$ and $\lambda$, we have evidence for our conjecture. In \cite{Creutzig:2013tja}
we have shown that the one-loop partition function of the higher spin theory can be reproduced by the 't Hooft limit of the coset \eqref{coset}, see also \cite{Candu:2013fta}. For the higher spin gauge theory, the one-loop partition function can be written in terms of a one-loop determinant, and the explicit expression may be found in \cite{Creutzig:2013tja} and references therein. For  the dual coset \eqref{coset} a state is labeled by $(\Lambda_{N+M};\Lambda_{N})$ in the 't Hooft limit, where $\Lambda_L$ represents the highest weight for SU$(L)$. In order to determine the spectrum of the theory, we need to specify how to take pairs of holomorphic and anti-holomorphic parts. Here we take a diagonal modular invariant as
\begin{align}
 \mathcal{H} = \bigoplus_{\Lambda_{N+M},\Lambda_N}
   \mathcal{H}_{(\Lambda_{N+M};\Lambda_N)} \otimes
   \bar{ \mathcal{H} }_{(\Lambda_{N+M};\Lambda_N)^*}
\end{align}
where the charge conjugated states are paired. Using the methods developed in \cite{Gaberdiel:2011zw,Candu:2012jq}, we can show the match of the partition functions in the 't Hooft limit once we assume the decoupling of so called ``light states.''
We can also show that the symmetry algebra matches for first few spins explicitly.

We may now be able to say that the duality between the 3d Vasiliev theory with U$(M)$ CP factor and the coset \eqref{coset} is more or less concrete. Thus the next question would be how the duality relates to superstring theory. Let us first review the arguments by Gaberdiel and Gopakumar \cite{Gaberdiel:2013vva,Gaberdiel:2014cha}. They focused on the case with $M=2$. In this case the coset model coincides with the Wolf space model, which is known to possess a large $\mathcal{N}=4$ superconformal symmetry \cite{Spindel:1988sr,VanProeyen:1989me,Sevrin:1989ce}.%
\footnote{It was already suggested in \cite{Henneaux:2012ny} to use the Wolf space model for the construction of a higher spin holography with extended supersymmetry.} With the large supersymmetry, we can identify the target space of superstring theory involved as AdS$_3 \times$S$^3 \times$S$^3 \times$S$^1$. However, now the higher spin theory includes only $2 \times 2$ matrix valued fields and it is not obvious how to see the relation to string states since the situation is much different from the one in \cite{Chang:2012kt}. Recently, they examined their conjecture more closely when the radius of one of two $S^3$'s becomes very large \cite{Gaberdiel:2014cha}. In this case the dual CFT has a small  $\mathcal{N}=4$ superconformal symmetry, and there is a considerable amount of literature on the duality between the 2d CFT and superstring theory.

However, if we want to apply the picture obtained in \cite{Chang:2012kt}, it is better to keep $M$ generic. Therefore, we utilize the coset \eqref{coset} with generic $M$. As we saw in
section \ref{ABJ}, we need to assign a U$(M)$ invariant condition to the CP factor of higher spin fields. Thus, instead of the coset \eqref{coset}, we consider the following coset as \cite{Creutzig:2013tja,Candu:2013fta}
\begin{align}
  \frac{\text{su}(N+M)_k \oplus \text{so}(2NM)_1}{\text{su}(N)_{k+M} \oplus \text{su}(M)_{k+N} \oplus \text{u}(1)_\kappa} \, ,
  \label{KScoset}
\end{align}
which is a Kazama-Suzuki model with $\mathcal{N}=2$ superconformal symmetry \cite{Kazama:1988qp,Kazama:1988uz}.
The target space of superstring theory dual to the coset is of the form AdS$_3 \times$M$_7$, but the $\mathcal{N}=2$ superconformal symmetry is not enough to determine M$_7$. Therefore, we cannot identify which superstring theory is involved in our triality. However, we noticed that the  $\mathcal{N}=2$ supersymmetry of the coset \eqref{KScoset} is enhanced to $\mathcal{N}=3$ at a specific value of the level $k=N+M$
\cite{Creutzig:2014ula}. With the extended supersymmetry the candidates of M$_7$ are quite restricted, and it is expected that we can construct an AdS$_3$ version of the ABJ triality by making use of the critical level coset model.

\section{Relations to superstring theory}
\label{N=3}

Let us first look for a 3d Vasiliev theory with $\mathcal{N}= p > 2$ supersymmetry. It is known to be difficult to extend the 3d Vasiliev theory to have extended supersymmetry with generic parameter $\lambda$. However for $\lambda=1/2$ the supersymmetry can be enhanced to a generic $\mathcal{N}= p > 2$ \cite{Prokushkin:1998bq,Henneaux:2012ny}. At this value of the parameter, the matter become massless and conformally coupled to gravity, and we consistently truncate the field content by half. We consider the case with $\mathcal{N}=2n+1$ $(n =0,1,2,\ldots)$,%
\footnote{See \cite{Candu:2014yva} for the cases with $\mathcal{N}=2n$.}
 whose supersymmetry algebra so$(2n+1|2)$ is generated by
\begin{align}
 T_{\alpha  \beta} = \{ y_\alpha , y_\beta \} \, , \quad
 Q^I_\alpha = y_\alpha \otimes \phi^I \, , \quad
 M^{IJ} = [\phi^I , \phi^J] \, .
\end{align}
Here we have introduced two types of parameters $y_\alpha$ $(\alpha =1,2)$ and $\phi^I$ $(I=1,2,\ldots , 2n+1)$ with the properties
\begin{align}
 [y_\alpha , y_\beta] = 2 i \epsilon_{\alpha  \beta} \, , \quad
 \{ \phi^I , \phi^J \} = 2 \delta^{IJ} \, .
\end{align}
The twister variables $y_\alpha$ organize fields with higher spin in a neat way.
Moreover, $\phi^I$ generates the Clifford algebra, which can be realized by $2^n \times 2^n$ matrices. The fields now depend on $\phi^I$, or in other words they are associated with a U$(2^n)$ CP factor.

The proposal here is that the higher spin theory with extended supersymmetry is dual to the coset \eqref{coset} with $M =2^{n-1}$ and a specific value of the level $k=N-M$ (after assuming some decoupling of free fermions) \cite{Creutzig:2014ula}. We assume the following non-standard Hilbert space as
\begin{align}
 \mathcal{H} = \bigoplus_{\Lambda_{N+M}} \mathcal{H}_{\Lambda_{N+M}} \otimes \bar {\mathcal{H}
}_{\Lambda_{N+M}^*} \, , \quad
   \mathcal{H}_{\Lambda_{N+M}} = \bigoplus_{\Lambda_N \in \Omega } (\Lambda_{N+M}; \Lambda_N) \, .
   \label{non-diag}
\end{align}
At large $N$ the label $\Lambda_N$ can be represented by two Young diagrams $(\Lambda_N^l, \Lambda_N^r)$ and $\Omega$ means that the sum is taken over $\Lambda_N^l = (\Lambda_N^r)^t$. Here $t$ represents the transpose of the Young diagram. For $n=0$, the duality reduces to the one proposed in \cite{Beccaria:2013wqa}.
Developing the techniques used in \cite{Beccaria:2013wqa}, we can show that the partition function in the large $N$ limit (with the assumption of the decoupling of light states again) reproduces the one from the dual classical higher spin theory. This means that the spectrum matches between the proposed dual theories.

In order to understand the meaning of the Hilbert space in \eqref{non-diag}, we move to another expression by making use of the level-rank duality in \cite{Kazama:1988qp,Naculich:1997ic}. We considered the coset \eqref{coset} with $k=N-M$, but a level-rank dual expression is given by the coset with $k=N+M$. We should remark that the number of decoupled fermions is changed. See \cite{Creutzig:2014ula}  for more detailed explanation. In the level-rank dual expression, a su$(N+M)$ factor with the level $k=N+M$ appears in the numerator of the coset \eqref{coset}. A crucial point here is that the $\text{su}(N+M)_{N+M}$ factor has a realization in terms of free fermions in the adjoint representation of su$(N+M)$. With this realization, the Hilbert space is generated by the free fermions modulo the factors in the denominator of \eqref{coset} and the repeated fusions of adjoint fermions yield the states in the representations of the form $\Lambda \in \Omega$ as discussed in \cite{Beccaria:2013wqa}.
Going back to the original form before applying the level-rank duality, the Hilbert space becomes the one in \eqref{non-diag}.
The symmetry generators of the critical level coset can be constructed by the free fermions. For examples, spin one currents can be given by bi-linears of fermions, while spin 3/2 currents are written in terms of the product of three fermions. Explicit forms of these currents can be found in \cite{Creutzig:2014ula}.

In this way we found another type of duality between 3d higher spin theory and 2d coset type model. Let us try to figure out the superstring theory related to these dual models (assuming that it exists).  From the lesson obtained in section \ref{ABJ}, we have to deal with singlets in the sense of the CP factor of the higher spin fields. Here we set $M$ to be a generic positive integer.
 Then, it is natural to think of the Kazama-Suzuki model \eqref{KScoset} with the critical level $k=N+M$. The higher spin theory dual to the Kazama-Suzuki model includes fields of the $2M \times 2M$ matrix form, but with U($M$) invariant condition assigned, see also \cite{Candu:2013fta}. One of the main results obtained in \cite{Creutzig:2014ula} is that the critical level Kazama-Suzuki model has a $\mathcal{N}=3$ superconformal symmetry.
From the dual conformal symmetry, the target space of superstring theory is fixed to be of the form AdS$_3 \times$M$_7$, as mentioned above. For the cases with  $\mathcal{N}=3$ superconformal symmetry, the known explicit examples are M$_7 =$(S$^3 \times$S$^3 \times$S$^1)/\mathbb{Z}_2$ in \cite{Yamaguchi:1999gb} and M$_7 =$SO$(3)/$U$(1)$ or SO$(5)$/SO$(3)$ in \cite{Argurio:2000tg}. The BPS spectrum and marginal deformations are studied in \cite{Argurio:2000xm} for the latter two models, and their result is consistent with those for our coset \cite{Creutzig:2014ula}. Therefore, we may conjecture that superstring theory on  AdS$_3 \times$M$_7$ with M$_7 =$SO$(3)/$U$(1)$ or SO$(5)$/SO$(3)$ and our coset are dual to each other. In order to examine whether this conjecture is true or not, we need to investigate our proposed triality in further detail.

\section{Conclusion}
\label{conclusion}

In this note we have reviewed our works on a lower dimensional analogue of ABJ triality in \cite{Chang:2012kt}. Extending the duality by Klebanov and Polyakov in \cite{Klebanov:2002ja}, the authors in \cite{Chang:2012kt} proposed a triality between 4d extended Vasiliev theory, superstring theory and the ABJ theory. We have explained why the extension of Vasiliev theory with CP factor is important to see relations to superstring theory. Inspired by the work, we have extended the duality by Gaberdiel and Gopakumar in
\cite{Gaberdiel:2010pz} such that the 3d extended Vasiliev theory with U$(M)$ CP factor in \cite{Prokushkin:1998bq} is involved. Our conjecture in \cite{Creutzig:2013tja} is that the dual theory is given by the coset model in \eqref{coset}.  We gave several supporting arguments for the duality, for instance, by showing the match of one-loop partition functions.
In order to see relations to superstring theory, we extend the duality to have more supersymmetry. In \cite{Creutzig:2014ula} we proposed a duality between 3d Vasiliev theory with extended supersymmetry and the coset \eqref{coset} at a critical level. Based on the duality we proposed that the Kazama-Suzuki model \eqref{KScoset} with the critical level $k=N+M$ is dual to a superstring theory with the help of $\mathcal{N}=3$ superconformal symmetry of the critical level model.

We have worked on a lower dimensional version since it is expected to allow to study the triality in more detail than the original ABJ triality. Indeed the 2d coset type models in \eqref{coset} and \eqref{KScoset} can be solved exactly, in principle. Moreover, the gauge sector of 3d Vasiliev theory is topological and dynamical degrees of freedom exist only in the matter sector. However, at least for our case, the supersymmetry is not so large to fix the dual superstring theory uniquely. We are currently working to make our conjecture more concrete, and we would like to report on our findings in the near future.
Recently, we started to understand the nature of marginal deformations of the 2d coset models. Higher spin gauge theory should correspond to the tensionless limit of superstring theory, so we need to deform the coset model to compare with superstring theory at a typical point of the moduli space. The higher spin symmetry is generically broken by the marginal deformation of the critical level Kazama-Suzuki model and the mass of higher spin fields generated through the breaking of higher spin symmetry is computed \cite{HR15, CH15}.


\providecommand{\bysame}{\leavevmode\hbox to3em{\hrulefill}\thinspace}
\providecommand{\MR}{\relax\ifhmode\unskip\space\fi MR }
\providecommand{\MRhref}[2]{%
  \href{http://www.ams.org/mathscinet-getitem?mr=#1}{#2}
}
\providecommand{\href}[2]{#2}

\end{document}